# A Survey on Feasibility and Suitability of Blockchain Techniques for the E-Voting Systems

**Umut Can Çabuk[1], Eylül Adıgüzel[2], Enis Karaarslan[2*]**

Dept. of Electrical-Electronics Engineering, Erzincan University, Erzincan, Turkey[1]

Dept. of Computer Engineering, Muğla Sıtkı Koçman University, Muğla, Turkey[2]

* Correspondence: `enis.karaarslan@mu.edu.tr`

**Abstract**: In the second decade of the 21st century, blockchain definitely became one of the most trending computational technologies. This research aims to question the feasibility and suitability of using blockchain technology within e-voting systems, regarding both technical and non-technical aspects. In today's world, although the course of this spreading is considerably slow, several countries already use means of e-voting due to many social and economic reasons, which we further investigated. Nevertheless, the number of countries offering various e-government solutions, apart from e-voting, is significantly high. E-voting systems, naturally, require much more attention and assurance regarding potential security and anonymity issues, since voting is one of the few extremely critical governmental processes. Nevertheless, e-voting is not purely a governmental service, but many companies and non-profit organizations would benefit the cost-efficiency, scalability, remote accessibility, and ease of use that it provides. Blockchain technology is claimed to be able to address some, obviously not all, important security concerns, including anonymity, confidentiality, integrity, and non-repudiation. The analysis results presented in this article mostly confirm these claims.

## I. INTRODUCTION

In the 21st century, the big boom of broadband internet technologies let e-governance and e-government be two interrelated rising trends. Many government-related services, such as institutional applications, information requests, document collection, tax and utility payments, legal issues and administrative operations can now be done online through special websites and mobile applications, in many countries. Likewise, similar business operations are also being moved to the online portals. e-voting, a considerably new concept that allows voters (citizens and members of businesses) to vote online through specific web portals and mobile applications, on the other hand, is still not widespread. It eliminates the need for distributed vote centers, paper ballots, ballot boxes and observer personnel; hence lowers the costs significantly. However, yet it is still rarely being used as the primary way of collecting people's choices and opinions regardless of the subject. E-Voting is principally different than online polls and questionnaires that are currently in use, since it involves %100 accuracy on real physical level (not account level) identity authentication for attending persons. Plus, confidentiality of both votes and voters is extremely important. Integrity and non-repudiation should be considered, too. Other social factors, like people's access to the Internet, socioeconomic costs, peoples' feeling of trust etc. cannot be neglected and are still arguable.

Recent major technical challenges regarding e-voting systems include, but not limited to [1]:

- Secure digital identity management: Any potential voter should have been enrolled to the voting system prior to the elections. Their information should be in a digitally processable format. Besides, their identity information should be kept private in any involving database.
- Anonymous vote-casting: Each vote, may or may not contain any choice/candidate, should be anonymous to everyone including the system administrators, after the vote is submitted through the system.
- Individualized ballot processes: How a vote will be represented in the involving web applications or databases is still an open discussion. While a cleartext message is the worst idea, a hashed token can be used to provide anonymity and integrity. Meanwhile, the vote should be non-repudiable, which cannot be guaranteed by the token solution.
- Ballot casting verifiability by (and only by) the voter: The voter should be able to see and verify his/her own vote, after he/she submitted the vote. This is important to achieve in order to prevent, or at least to notice, any potential malicious activity. This countermeasure, apart from providing means of non-repudiation, will surely boost the feeling of trust of the voters.

These problems are partially addressed in some recent applications. Yet, means of e-voting is currently in use in several countries including Brazil, United Kingdom, Japan, and Estonia. Estonia should be evaluated differently than the others, since they provide a full e-voting solution that is, said to be, equivalent of traditional paper-based elections. We will briefly present current applications, previous attempts, and future plans in the next section already, however it is good to state some common problems of current e-voting applications here [1];

- High initial setup costs: Though sustaining and maintaining online voting systems is much cheaper than traditional elections, initial deployments might be expensive, especially for businesses.
- Increasing security problems: Cyberattacks pose a great threat to the public polls. No one would accept the responsibility if any hacking attempt succeeds during an election. The DDoS attacks are well known and mostly not the case in the elections. The voter integrity commission of the United States gave a testimony about the state of the e-elections in the US recently. Accordingly; Ronald Rivest stated that "hackers have myriad ways of attacking voting machines". As an example; barcodes on ballots and smartphones in voting locations can be used in the hacking process. Appel stated that we mustn't ignore the fact that computers are hackable, and the evidences can easily be deleted. Double-voting or voters from the other regions are some common problems [2]. To mitigate these threats, software mechanisms which promise the following should be deployed:
• Prevention of evidence deletion.
• Transparency with privacy.
- Lack of transparency and trust: How can people surely trust the results, when everything is done online? Perceptual problems cannot be ignored.
- Voting delays or inefficiencies related to remote/absentee voting: Timing is very important in voting schemes; technical capabilities and the infrastructures should be reliable and run at the highest possible performance to let remote voting be synchronous.

The blockchain technology may address many issues regarding e-voting schemes and make e-voting cheaper, easier, and much more secure to implement. It is a considerably new paradigm that can help to form decentralized systems, which assure the data integrity, availability, and fault tolerance. Some state that "the blockchain technology is bringing us the Internet of value: a new, distributed platform that can help us reshape the world of business and transform the old order of human affairs for the better." [1]. This technology aims to revolutionize the systems. The blockchain systems are formed as decentralized networked systems of computers, which are used for validating and recording the pure-online transactions. They also constitute ledgers, where digital data is tied to each other, called the blockchain. The records on the blockchains are essentially immutable.

## II. RELATED WORKS

The e-voting concept and the blockchain technology share a common fate, that is, both of them started to be used publicly and commercially, before the academic studies pave the way for better solutions and standardization. Due to lack of academic collaborations, their development has been conducted predominantly out of the academia, until recently. For example; a promising collaborative project mostly funded by the European Union, called "An innovative cyber voting system for Internet terminals and mobile phones" was ended prematurely, without concluding with a voting system as a product [3]. Hence, in this section we preferred to focus more on public and commercial advances, without ignoring researches.

*A. E-Voting Applications*

Your paper must use a page size corresponding to A4 which is 210mm (8.27") wide and 297mm (11.69") long. The Since 1990s, e-voting, at least as an idea, is being placed in laws and regulations in many countries. But, only a few of these countries has ever used a real implementation during official elections or referenda. And even fewer of them are still using e-voting. Two main reasons we have noticed of quitting use of e-voting are broad suspicion of frauds and implementational/deployment costs. We have looked through some of the remarkable examples throughout the world. However, the e-voting attempts are surely not limited to these, especially the voting machines were used widely, and are still in use in some states. They are out of our scope and not comprehensively included in our survey. More information on them and their security flaws, can be found elsewhere [4].

The United Kingdom for example, despite the increasing requests from the people, did not yet offer any means of online voting, mainly because of the possibility of cyberattacks that aim to disrupt the election results. Threats regarding anonymity of ballots, identity checks and duplicated votes are also big concerns. Studies on finding a secure solution are yet being conducted, through [5].

In 2012, authorities in France allowed their citizens who live abroad to vote online through a web portal that has been developed by Scytl SA, a private company founded as a university spin-off [6]. The online voting processes has been later cancelled by National Cybersecurity Agency due to increasing risks of cyber-attacks [7].

Means of electronic voting was also banned in the Netherlands in 2007; and in 2017, electronic vote counting was abandoned, too. Turning back to paper ballots and manual counting is reportedly done because of security concerns, namely, preventing foreign manipulation in the elections [8].

In Brazil, voting machines are in use since 1996. These machines, also called kiosks, are modified computers that run specially designed software, which stores and counts the number of votes entered by the attached keypad. There are strong debates regarding the security of these devices, since the source codes of the embedded software are not open to the public. Some authorities claim that the devices are secure as long as the internal software is proper; but if the software is hacked via external access, which may not be hard, then it would be a disaster [9].

Estonia, on the other hand, provides a full e-voting solution for the governmental elections [10]. Unlike efforts in other countries, their implementation is persistent and more comprehensive, also more reliable. There might be some technical/systematic drawbacks, however the government's intention regarding building an e-country is impressive. As a side note, Estonia government also issues e-residency, a form of subscription to some online services, including business activities. Every citizen in Estonia possesses an ID card equipped with a digital chip that contains identity and biometric information. People who want to attend online elections, which are held in parallel with conventional elections made in vote centers, should use their ID card and a card reader (provided by the authorities) to login the specially designed web portal or e-voting application, called i-Voting. It is also possible to use a mobile ID, that involves an SMS authentication, in later versions. The i-Voting application can be reached via its web portal, www.valimised.ee, during the elections period. It remains active for several days, until the traditional election's day (which is only 1 day). The ID card acts as a token (possessive security factor) and the voter should also type a unique predefined password (knowledge factor) for authentication himself/herself to the system [11].

In Estonia, another similar system is also implemented for a slightly different purpose. The parliament of Estonia, in their dedicated official web portal, namely www.rahvaalgatus.ee, allows their citizens to start online petitions, make inquiries and propose new acts/laws. Moreover, people may seek, read, and digitally sign other peoples' proposals, if they like or agree. When a petition reaches 1000 signatures, it is later discussed in a special session in the parliament.

In another previous work, the researchers have proposed an inclusive e-Democracy model, where e-voting systems are used by both institutions and people to improve the democracy by allowing people to create/suggest administrative polls and attend official elections online [12].

*B. Blockchain-based Projects*

The blockchain technology owes its vast popularity to Bitcoin, the very first cryptocurrency. Shortly after, new cryptocurrencies arise. However, possible uses of blockchain is surely not limited to digital currencies. By April 2014, it is known that this chain is used for more than 80 different purposes under the name of sidechain [1]. Since 2014, the term Blockchain 2.0 has been used for decentralized and effective registry implementation and innovations. It is important to note that; in addition to Bitcoin's monetary value and its usage in online trade, it has shown a prototype of a reliable distributed database system. This property of the blockchain let many to think about other usages; such as fintech, implying use of blockchain in the finance sector and its digital applications. E-Voting is, or might be, another theme for using blockchain techniques [13]. Since there are already several projects regarding that, we do not dig much into cryptocurrencies, instead will introduce these e-voting studies.

Researchers from Kyoto University, proposed a novel e-voting scheme, in their 2016 technical report [14]. Though they don't mention any implementation, they provided a comprehensive model, which suggests a way of exploiting the coin transfer mechanism to achieve a "vote transfer" mechanism instead. Per to the model, all the registered voters (as well as candidates) are issued an official coin wallet. The electoral administration provides coins to be used for the voting process, to the voters. Voters then, briefly, transfer their coin(s) during the election day(s) to the candidate(s) of their choice, just like a regular coin trade/transfer. Yet there is also an intermediate unit, located between the voters (senders) and the candidates (receivers), that converts the transferred coins to another coin-currency and resumes/completes the transfer. So that, it is no more possible to trace the owner of any vote. This is done to protect privacy and anonymity of the voters. The researchers have used Bitcoin and Zerocoin for this purpose, however these can be replaced by others in a real implementation, depending on the use case scenarios.

Blockchain technology, on the other hand, has some technical barriers to solve in order to become a large-scale and popular solution [15]. The technology needs some more time for its maturity. It should also be noted that blockchain is not a perfect solution for all the problems, this concept will be covered in the feasibility analysis section.

*1) Implementations and Design Decisions.*

Blockchain technology can be integrated into current e-voting schemes as [14] suggests. Though the idea is quite new, there are also several active implementational projects those worth analyzing. Selected projects and required ledger environments are investigated and some of their methodologies, pros and cons are given. One of them are further examined and how it works is explained later. The projects are classified according to the following characteristics, some of which are listed in Table 1:

• Proof of Identity: The identity of the user should be validated to use the system which is one of the most important security issues. This process involves using ID cards, or digital residence code or access to the official digital profile. The user must broadcast a proof of his/her identity when using Sovereign. The user should satisfy some criteria that can only be met by human judgment to avoid an AI from interfering with the process. Some necessities include "film proof", "hash proof", "validate proof" which are specified in the products whitepaper (github.com/DemocracyEarth/paper). Polys requires identification with unique code and mail verification combination. Boule has the biometrical recognition feature. StakeWeightedVoting stipulates picture and government id card along with ID key to the voting booth.

• User Interface: User interface is mostly web based in the e-voting platforms. Admin panel has included creation of vote, tracking process. User panel is more restricted, after verification it gives only permission to vote and results. Polys gives the organizer panel to whom that created the poll. In Sovereign there is no difference between admin and user interface. When you signed up you can create a vote and you became an admin of that voting process or just vote as a user. StakeWeightedVoting uses Graphene as user panel.

• Token: A token can be defined as a type of digital asset created and presented within the scope of a service or a platform application and it used to fulfill all the functions of that service. The crypto-currency units such as Bitcoin (BTC) and Ether (ETH), which we are now familiar with, are one token, but each token does not have to be a crypto-currency. While Sovereign using Bitcoin as token, Boule uses BOU Token which is not a solicitation for investment. BOU Token usage is restricted by rules which are specified in the terms (www.boule.one/terms.html). The token is designed only for particular uses with respect to the Boulé ecosystem, it is not necessarily merchantable and does not necessarily have any other use or value. Vote organizers and institutions will buy the service using BOU Token. It will be distributed to the voters to allow them to access the ballot and cast their vote. StakeWeightedVoting is using BitShares (BTS) token. Using web-based version of the Bitshares wallet. The user should navigate to settings and then he/she would import the wallet. Once the user has imported the Bitshares keys into this new wallet to user interface, admin should be able to navigate to user's account and see all of the things user would normally see when working with the Bitshares wallet interface.

TABLE I
CLASSIFICATION OF THE SELECTED BLOCKCHAIN-BASED VOTING PROJECTS

|  | *Ledger* | *Language* | *Token* | *Consensus Protocol* |
|---|---|---|---|---|
| ***StakeWeighted Voting*** | BitShares Blockchain | C++ | Bitshares BTS | Delegated PoS (DPoS) |
| ***Polys.me*** | Ethereum | Solidity | Unspecified | PoW |
| ***Boulé*** | Ethereum | Solidity | Boulé BOU Token | PoS |
| ***Sovereign*** | Bitcoin | Python | Bitcoin BTC | PoW |

Ledger: Ledger is a type of database that is shared, replicated, and synchronized among the members of a network. The ledger is distributed and records the transactions, such as the exchange of assets or data, among the participants in the network [16]. StakeWeightedVoting depends on the BitShares chain, which is being used for years and being developed by the open source community. This may show that it is open for advancement and is more prepared for the attacks. Polys and Boulé is using Ethereum, which is also a stable, strong, and trustworthy blockchain as reported. Sovereign declares Bitcoin as the most popular and reliable ledger and uses it.

• Consensus Protocol: The nodes, which share public ledgers need to use consensus algorithms to agree on a decision. The algorithms have to be functional, secure by design and also efficient. The consensus mechanisms are used by the nodes of the system to decide, mainly on who will get the right to update the block. This will ensure the integrity of the data recorded on the blockchain. Bitcoin and other mining based crypto coins use Proof-of-work (PoW) which depends on every node using their GPU power for that. This process is slow and uses extensive electricity, so alternatives are being developed. Sovereign, Boule are using PoW algorithm as consensus protocol. The most common alternative is Proof-of-stake (PoS), which Ethereum is planning to use. The 'validator' uses stakes and should invest in the coins of the system in PoS. Polys depends on PoS algorithm. StakeWeightedVoting is using a variation of PoS, which is called Delegated PoS (DPoS) [17].

• Programming Languages: Many programming languages can be used to develop blockchain applications and smart contracts. A smart contract is a computer protocol intended to digitally facilitate, verify, or enforce the negotiation or performance of a contract. Smart contracts allow the performance of credible transactions, which are trackable and irreversible, without intervention of any third parties (i.e. software or people) [18]. C++ and Python are widely used. Ethereum based solutions tend to use Solidity language for implementing smart contracts. Sovereign was written by Python, but uses JavaScript based Meteor environment as its web-based user interface.

• Open Source: The users will trust the systems when they see the code. StakeWeightedVoting and Sovereign are open source. Users can install and even contribute the source code. StakeWeightedVoting has step-by-step guide for Ubuntu terminal installation. Sovereign has both Ubuntu and Mac OS terminal installation guide and a Windows installer. Users purchase the web-based service they provided, from Polys and Boule without any installation file.

*2) A Case Study: StakeWeightedVoting.*

StakeWeightedVoting (github.com/FollowMyVote/StakeWeightedVoting), is chosen as the project-of-interest for our case study among others, because of its being open source, documentation, GUI, and the chance of making change on it. Mainly, this application can be used to vote for one more proposals (candidates) by sending (digital) coins as vote ballots, from voters to candidates of their choice. Votes here can be weighted proportional to the voters' balance of BTS coins in their unique wallets. This might be useful for some polls and contests, rather than elections. As claimed, unique wallets are created in association with the users' (voters) accounts. Yet, supposedly, there is no connection between the voters and their votes, despite this association. As an important note, this application is designed to run on voting booths, special purpose computers located in vote centers, instead of personal computers of voters. The workflow of the system is explained step by step, below:

1. The voter (user) submits his/her picture and the government ID card, along with his/her ID key (a public key) to the voting booth.
2. ID (identity) verifier, receives this personal info of the voter and approves the ID key, by consulting its user database. The ID verifier certifies on the blockchain that the given ID key is unique, which authorizes the ID key (of the voter) to vote on a certain ballot type. Access to a certain ballot type is later granted by the ID verifier.
3. The certification from the ID verifier is paired with a blinded token (briefly an encrypted digital message with no specific author) and sent to the registrar, requesting a signature of approval to request a certain ballot type.
4. The registrar signs the blind token (without seeing the content of the message) and sends it back to the voting booth. Hence proves its authenticity.
5. The voting booth sends a signed un-blinded token with a vote key to the registrar, requesting certain ballot type.
6. The registrar certifies the voting key for a certain ballot type.
7. If certified by both the ID verifier and the registrar, the voter has now been authorized to cast a ballot. Using the vote key, the voting booth user interface generates a ballot for the voter to cast their vote.

Use of a blind signature mechanism cryptographically enables to cast fully private ballots with non-revealable owners. If not used, it might be possible to find the voter of any vote by investigating the blockchain. Because all transactions, especially the coin (vote) transfers, are normally written to the chain.

### III. FEASIBILITY ANALYSIS

This section provides results of our survey regarding the feasibility of taking advantage of blockchain technology in e-voting applications. By speaking of feasibility, we mean a cost-effective, scalable, secure, and easy-to-deploy system (or subsystem). While it is hard to determine global measures of these properties, we have considered some thresholds or factors in order to determine whether it is feasible to replace the existing (and prospective) structures with their blockchain-based counterparts, or not. Any blockchain-based solution should be (noticeably) cheaper than the traditional elections in the long term, say in a 3-years period with at least 1 elections per year [19]. It should not be more expensive than non-blockchain solutions either. The system should support millions of people, depending on the country, business size or target group population. The security level should not be lower than non-blockchain solutions, details regarding security requirements are given later in the text.

*A. Blockchain Fundamentals*

Blockchain, by its nature, cannot be applied to all systems as a modular off-the-shelf solution. The feasibility flowchart that explains when a blockchain database is useful, is shown in the Figure 1. This diagram is a simplified version of the flowchart, which is given in [20]. If blockchain is not useful/suitable for a project, it cannot be feasible, too. Blockchain solutions are suitable, when the following characteristics are present in the legacy subject systems [1]:

- Shared data: When there is structured information which should be shared between entities,
- Multiple parties: When there is a need for more than one entity which reads or writes the data,
- Low trust: When there is no presumed full trust between the members of the system,
- No trusted third party: If it is not available or not preferred due to implementation difficulties or costs,
- Auditability: If we want the records to be immutable (not to be changed or deleted after recording).

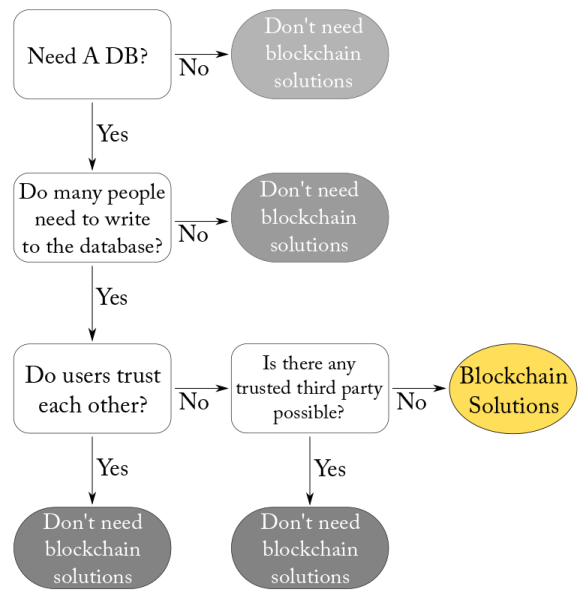

Fig. 1. Blockchain solutions feasibility flowchart.

Blockchain-based systems may also have the following value-added characteristics [1]. These values can make the blockchain-based alternatives preferable (also feasible) depending on the use case scenarios.

- Disintermediation: The transactions are not verified by a single central gatekeeper and this may reduce the costs of building an infrastructure and its maintenance, and there may be some performance gains.
- Transaction interaction: Complicated and interconnected transactions can be implemented by using smart contracts. Blockchain structure provides an easy and modular base for public key infrastructures and blind signatures.
- Auditability: Every record in blockchain keeps who are involved in the transactions, as well as the type, amount the content.

Blockchain based systems can provide full trust, so that no one is in the center of the process as an administrator, and complete privacy for all parties when properly designed. Blockchain is a candidate for e-voting as it provides stability, fault tolerance and singularity features of the democratic elections. Moreover, it completely enables the secret ballot and open counting policy, just like traditional paper-based elections. A properly designed system should aim to provide the following fundamentals [1]:

- Full trust, such that the election should not be under the control of anybody. It should be guaranteed that the election results can't be manipulated and there will not be inconsistency between the records in different intermediate systems, if any.
- More transparency than the legacy online systems (as well as traditional elections). Most blockchain implementations allow listing all the transactions made, with a content and a timestamp, but without revealing the parties who are involved in the transaction. It is also possible to hide the content for a while. Thus; all casted ballots can be listed in real-time, and all votes can be counted by any of the observers, while protecting the privacy of the voters. Additionally, the voters have the opportunity to verify their votes, after they casted their ballots.
- Cost savings, due to diminished needs for expensive servers or computers and open source software.
- Remote participation, also called as online absentee voting. If desired, it will help to increase the participation rates and will make the attendance cheaper for voters. That would be preferable especially for businesses and non-profit organizations.
- Immutable and non-manipulatable records of votes. Once the election is over, it should technically not be possible to manipulate the votes, even by the system technicians nor administrators. The opportunity of updating votes (only legal and valid changes) could be provided during the election time. This should be implemented with a supporting consensus protocol. Security is further discussed in the following parts.

*B. Social Aspects*

Applications involving e-voting and the Blockchain technology have big social impacts, too. These impacts can further be listed as the value obtained from the provided ease of use and the people's perception of trust to these so-called "hi-tech" systems. Generally, the e-Government services enabled wider, easier, and faster access to the government services for the people, including the ones living in remote settlements and the ones who are very busy and/or mobile. So that it can be seen as a powerful tool that reinforces the government-citizen relationships [21]. While the e-Government itself is not directly related to the democracy, the concept of e-voting extends the e-Government to provide means of democracy, called e-Democracy.

The ease of use and the financial benefits of such e-services are no more under discussion [22], but the perception of trust, a newer concern brought by the services related to e-democracy, might be shadowing these benefits, when it comes to e-voting. Independent from the topic and theme, if the majority of the voters do not trust the proposed e-voting system, then this system should not be accepted as the only way of voting. This applies even if the concerns are totally void and unsound or just conspiracy. This also explains why Estonia still holds both traditional and online elections together. Several studies showed that the trust in such electronic voting systems are considerably low (or at least not high as the traditional systems) [23], and some institutions and researchers started to propose ways to improve that trust already [24-25]. The trust in e-voting, if can be increased successfully, would even increase the overall trust to the general political system, especially in the developing countries [26]. Use of blockchain technology, which is used in the popular cryptocurrency Bitcoin (and many others) may strengthen the perception of trust, since Bitcoin and other cryptocurrency transactions are widely known to supply trust to the transactions, even between untrusted parties, as long as users are aware of some security countermeasures. If the software is open source, (better with free software license), public opinion and trust will even be higher. Because, open source codes can be investigated by everyone who wants to contribute to the project (like in Estonia). In such a case, even malevolent people unintentionally help developing the system.

*C. Financial Aspects*

Undoubtedly, using automated electronic systems, including web portals and mobile applications, will lower the administrative costs in the long term, despite their higher initial investment costs [21]. A previous study showing a rough comparison regarding infrastructure and maintenance costs of traditional and electronic elections was recently made [19]. Per to the study, the advantages of switching to an online elections system may provide savings up to several times per year. The difference becomes sharper, especially if there are 2 or more elections throughout a year. The cost from the point of view of the voters is analyzed in another work in Estonia [27]. This study briefly suggests that the voters who live at least 30 minutes distance from their voting centers have higher (time and money) costs, and thus certainly are more likely to prefer voting online. But, elderly people's unwillingness to use computers is neglected, as they noted.

Cost of traditional elections mainly consist of material (ballot papers, boxes etc.), personnel, and logistics costs. Yet the cost of e-voting systems includes software development, hardware infrastructure and related maintenance costs. Use of blockchain based solutions may even lower these software costs, since most blockchain bundles are open-source projects and come with customizable APIs. Besides, integrating the blockchain-based e-voting system with some sort of cryptocurrency-payment system may provide different setups and opportunities.

*D. Security and Reliability*

The security services that the blockchain provides is compared with other database solutions in Table 2. The availability and the fault tolerance of the system is high as all the nodes keep a copy of the records and check each other to make a stable system. The blockchain provides transparency with anonymity. The privacy is not aimed but can be implemented.

TABLE III
COMPARISON OF THE SECURITY SERVICES OF DIFFERENT SOLUTIONS [31]

|  | *Blockchain* | *Database* | *Distributed Database* |
|---|---|---|---|
| *Integrity of the Records* | High | Moderate | Moderate |
| *Availability* | High | Low | Moderate |
| *Fault Tolerance* | High | Low | Low |
| *Privacy* | Low | High | Moderate |

Each block keeps the hash of the previous block and this eventually provides a chain of blocks that are linked to each other as shown in Figure 2.

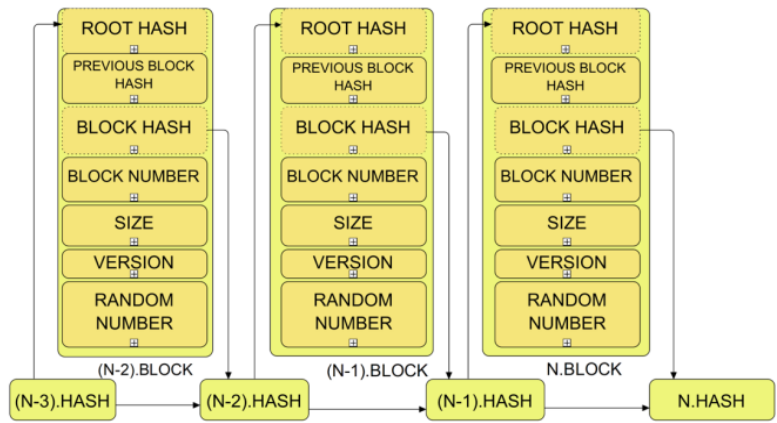

Fig. 2. Blockchain Structure [28].

Merkle tree is used in order to keep the integrity of the records. Its structure is shown in Figure 3. Each block holds more than one transaction. Firstly, hash values of each transaction are taken and paired with the hash of the other transaction. Pairs of hashes are then combined till it results in a single root hash. Transactions can easily be verified by this structure.

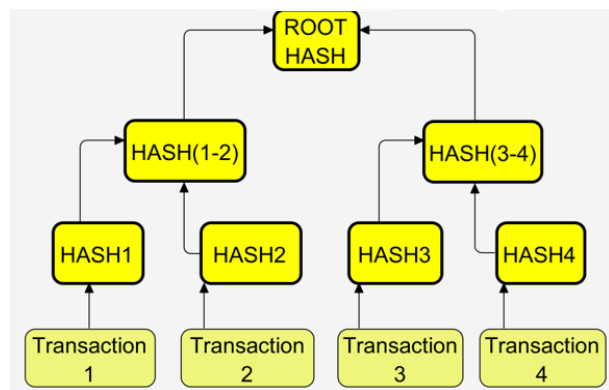

Fig. 3. Merkle Tree Structure [28].

Blockchain based systems are said to keep immutable records, but there is a tricky way to update the records (blocks), which is managed by the consensus protocols. Hence, the reliability of the system depends on the implemented consensus protocol. The consensus protocols (PoW, PoS, etc.) are the rules that determine which node will have the right to write to the blockchain. Bitcoin and mining-based blockchain implementations mainly use the PoW (Proof of Work) algorithm, which depends on the possessed computing power. Anyone who wants to change a block, should change that block and all of the following (next) blocks in the blockchain. For example, in a chain of 1000 blocks, if a user (or attacker) wants to change (only) the 100th block, then he/she should change all the blocks starting from 100th until 1000th. The attacker (node/computer) should get the writing turn for every block, and for PoW, this requires that it should have at least 51% of all of the computing power provided by (the sum of) all the nodes contributing to the chain. This attack is theoretically possible, especially in small networks, but as all the details of the transactions are recorded in all copies of the blockchain, it will be very easy to detect such a malicious activity; more, this negative effect can be mitigated in a short time [29], since it is not hard to exclude a node from the network.

The identity verification (authentication) is implemented by asymmetric cryptology in crypto coins. The public wallet address is the public key and private key is obtained from the public key. Elliptic curve cryptography (ECC) is widely used in this process. The keys are used to form Message Authentication Code (MAC) for signing the transactions. The MAC protects the integrity and authenticity of the message. These keys are also used to share the session key, which will be used to encrypt the communication for confidentiality. Additional encryption can be done for different applications. While this mechanism ensures the integrity and authentication of the user/voter accounts, authenticating the users themselves remains as an open problem.

*E. Comparison with Alternatives*

While there are hundreds of different e-Government applications and services, throughout the world, we cannot claim the same for e-voting systems, there aren't many to compare. Even so, we compared the distributed model of the blockchain scheme with centralized solutions, mostly referring to the Estonia model, which is one of the few successful applications so far.

Since the application in Estonia is a full off-the-shelf solution, a blockchain based variation shall not totally replace all the methods used in parts of that system. For example, a blockchain based solution itself would not provide a brand-new personal ID authentication mechanism, nor a different web portal. Likewise, they will share a similar blind signature mechanism, where the digital ballots are signed by an authoritative software using public-private key pairs, right after removing the ID of the voter, if it is ever somehow included.

The main differences are the computation and storage architectures. Classical relational databases are secured by their infrastructure, which may (or may not) include firewalls, access control and encryption. Blockchain databases, on the other hand, are secured by pure cryptology. This security implies encryption, integrity, confidentiality, transparency, and non-repudiation. Having confidentiality together with transparency is the true need of an election system. Literally; voters' identities must not have any kind of relation with their casted votes, duplicate voting must be prevented, voters should be able to verify their ballots, and the vote storage/counting phases should be time trackable. Providing all these features at the same time is very hard with the classical relational databases. Conversely, these can be implemented easily with a blockchain based system. In addition, a classical database is located in a single central point, which creates a single-point-of-failure. In case of a critical error, or an attack (hacking/hijacking attempt) it is very likely to completely lose the (control on the) database. Yet the blockchain based databases are always backed-up with many up-to-date copies, distributed in all (or some) of the computers in the network. Further; while the blockchain based databases have proven scalability, the classical databases may face (and cause) severe performance degradations, if the number of users and transactions become very high. By the way, it worth noting that the scalability of a blockchain system is highly implementation and consensus protocol dependent, the transaction rate may vary from 30 per second up to many more transactions. There is a speed-security tradeoff, and it is considered in academic studies [30] and new implementations.

*F. SWOT Analysis*

A blockchain-based e-voting system will have all the pros of e-voting systems (against traditional elections) but may not pose some of the cons of non-blockchain-based (naive) e-voting systems. Here we summarize some of our findings in Table 3. Our analysis showed that, the potential gains of holding blockchain-based online elections are significant and worth developing. However, there are also non-negligible concerns regarding the implementation details, use cases and extreme conditions.

## IV. CONCLUSION

The blockchain technology offers a decentralized storage and computation mechanism for e-voting systems, where the voting records are transparent to all the voters and independent observers (as long as they also pose an account or wallet etc.). It offers a system, in which everyone can trust. This trust is not just about the perception, but rather the mathematical, analytical, and logical means of security, provided by the blockchain technology. Per to our study, using blockchain mechanisms that supports smart contracts (or similar), such as Ethereum, can in fact be a good fit, since it would natively support distributed applications on the chain.

The aforementioned security term includes, integrity, verifiability, and non-repudiation of votes; authentication and singularity of voter accounts; immutability and trackability of all the records. Privacy and confidentiality of votes, is rather implementation dependent, yet very important to consider. Normally, it is possible to find the relation between a voter and his/her vote, by digging into the chain, if there is no unique prevention, like presented in [14].

TABLE IIIII
SWOT TABLE FOR BLOCKCHAIN-BASED E-VOTING, NAÏVE E-VOTING AND TRADITIONAL ELECTIONS

|  | *Blockchain-based e-Voting* | *Naive e-Voting Systems* | *Traditional Elections* |
|---|---|---|---|
| *Strengths* | • Immutable records. Record deletion is nearly impossible; even if it is successful, the evidence deletion can be prevented.<br>• Provides transparency with privacy.<br>• Cheaper in the long term.<br>• Enables elastic elections: variable durations, conditions, and target groups.<br>• Provides instant results. | • Cheaper in the long term.<br>• Enables elastic elections: customizable durations, conditions, and target groups.<br>• There are several sample models already. | • People trust the paper-based voting and counting, as long as the process is transparent.<br>• Does not rely on internet and computers, good for regions with low internet existence/usage. |
| *Weaknesses* | • The strengths will depend on the implementation.<br>• Technology is new and there are scalability issues. The performance may degrade on high usage.<br>• Inadequate tooling: Developer and testing tools are not enough [15]<br>• Initial deployment costs are salient, but lower than naive solution. | • Initial deployment costs are high.<br>• Perception of trust is low.<br>• Internal processes and casted votes are less transparent.<br>• Uses non-scalable classical databases.<br>• Existing unsuccessful attempts may disrupt the motivations. | • Costs are very high in the long term.<br>• In-person attendance may be hard and annoying.<br>• Physical security is though and expensive.<br>• Not possible to set vote centers in small and far-away settlements.<br>• Crowded vote centers become open targets for terrorism. |
| *Opportunities* | • New solutions to improve voting transparency<br>• Secure remote participation and voting<br>• Secure storage and records.<br>• Once learned, easier for elderly and disabled people<br>• Might bring more democracy to government units, local administrations.<br>• Less bureaucracy for businesses. | • Secure remote participation and voting.<br>• Once learned, easier for elderly and disabled people<br>• Might bring more democracy to government units, local administrations.<br>• Less bureaucracy for businesses. | • Less prone to conspiracy theories.<br>• Easier and cheaper for smaller and non-distributed groups. |
| *Threats* | • If the cryptographic keys are compromised, the attackers can abuse the system.<br>• Consensus protocol should be chosen and used wisely. There is a risk of a party who could dominate the P2P network by power in PoW, also by stake (wealth) in PoS algorithms. | • People's perception of trust is significantly lower.<br>• The centralized processing and storage architecture creates a single-point-of-failure.<br>• The centralized structure creates an easier target for attackers, too. | • Human-factor may cause errors during counting.<br>• Physical attacks may block or distort the voting process.<br>• Re-holding elections are extremely costly, in case of appeals.<br>• Difficulties regarding holding elections may result in having less elections. |

The features that this technology provides may seem perfect at the first glance, but we are not yet fully aware of all the risks regarding security and scalability, as blockchain-based e-voting systems are (mostly) still on the testing phase. As a result of our investigation we can say that the number of e-voting platforms are increasing day by day, and they all claim that they are more secure and less costly for voting. On the other hand, remote participation does not seem to be secure enough with the current technology, so we will still need the user to show-up in person (at least in most countries and businesses), as there are still risks regarding the personal ID authentication. As an example; the voters' devices might be compromised, and his/her cryptographic keys can be used maliciously. However, prevention of the evidence deletion and the possibility of reaching transparency with privacy are important and should further be studied in depth. Other further studies will include developing models on its usage and testing the currently under development blockchain based election systems by implementing security attacks. In parallel, novel ways of authenticating the voters themselves (not accounts), including the biometric factors, should be researched.

## BIOGRAPHIES

**Umut Can Çabuk** holds a BSc degree in Electronics Engineering (2012) from Uludag University, Turkey, and a MSc degree in Information Technologies Engineering (2015) from Aarhus University, Denmark. He is working as a research assistant in the telecommunications unit of electronics engineering department in Erzincan University, Turkey since 2016, and currently studying his PhD in Information Technologies in Ege University, Turkey. His research interests include wireless technologies, mobile networks, sensor networks, Internet of Things, information security and cryptography.

**Eylül Adigüzel** is a MSKU CENG Blockchain Research Group member, who is eager to learn the fundamentals and technical barriers of this technology. She is working on using blockchain effectively for e-government systems for her finishing thesis. She is a senior student in MSKU Computer Engineering department.

**Enis Karaarslan** is an Assistant Professor of the Department of Computer Engineering of Mugla Sitki Kocman University. He received the B.S. in Computer Engineering (1998), the M.Sc. in Computer Science in 2001, and Ph.D. in Computer Engineering (2008) all from Ege University. He had been a post-doc researcher in EC JRC-IPSC (Institute for the Protection and Security of the Citizen), Italy, from 2011 to 2012. He established MSKU NetSecLab for network/security education and research. His research areas are computer networks, security, privacy and blockchain. He has over 30 papers to his name.